\documentclass{article}

\PassOptionsToPackage{numbers, compress}{natbib}
\usepackage[preprint]{neurips_2026}

\usepackage[utf8]{inputenc} 
\usepackage[T1]{fontenc}    
\usepackage{hyperref}       
\usepackage{url}            
\usepackage{booktabs}       
\usepackage{amsfonts}       
\usepackage{nicefrac}       
\usepackage{microtype}      
\usepackage{xcolor}         
\usepackage{amsmath}
\usepackage{graphicx} 
\usepackage{amssymb}
\usepackage{multirow}
\usepackage{listings}
\usepackage{xcolor}

\lstdefinestyle{promptstyle}{
    basicstyle=\ttfamily\small,
    breaklines=true,
    breakatwhitespace=false,
    columns=fullflexible,
    keepspaces=true,
    frame=single,
    framesep=5pt,
    xleftmargin=5pt,
    xrightmargin=5pt
}

\title{PSD: Pseudo Self-Distillation of Memory Representation Capabilities for LLM Agents}

\author{P. Suhail, Menglin Xia, Xuchao Zhang, Mayukh Das, Chetan Bansal, Saravan Rajmohan \\
    M365\\
    Microsoft Research\\
    }

\begin{document}

\maketitle

\begin{abstract}
Memory systems are becoming a core component of LLM agents, but constructing and maintaining memory remains expensive because it relies on repeated calls to large proprietary language models. This cost creates a major barrier to deploying memory-enhanced agents at scale. In this paper, we present Pseudo Self-Distillation (PSD), a framework that enables small language models (SLMs) to construct hierarchical memory representations by distilling behavior from a strong black-box oracle through a multi-stage training pipeline. Standard distillation methods require access to teacher logits or hidden states, which closed models do not expose. Unlike conventional self-distillation settings, where supervision is derived from a model's own predictions, sampled rollouts, or aggregated outputs, PSD enables a single-model distillation setup while channeling external oracle knowledge through the prompt. PSD uses a single small model in two roles: a teacher that sees a privileged prompt containing the oracle's answer as reference context, and a student that sees only the task prompt. The student learns to reproduce the teacher's output distribution, absorbing oracle-guided behavior into its own weights without accessing the oracle's internals. On LoCoMo, PSD-trained Qwen3-0.6B, 1.7B, and 4B match or exceed GPT-4.1-mini on downstream retrieval at a fraction of the deployment cost, with off-policy PSD achieving the strongest results across most conditions. We further show that this memory-construction capability transfers out-of-distribution to LongMemEval, despite the students being trained exclusively on LoCoMo with no exposure to LongMemEval data.
\end{abstract}

\section{Introduction}

Large language models (LLMs) have enabled increasingly capable agents that can plan, reason, and interact over extended periods of time. However, effective long-horizon behavior requires more than strong reasoning capabilities. Agents must retain, organize, and reuse information accumulated across many interactions. As a result, memory has emerged as a fundamental component of modern agent architectures. Recent work has explored a wide range of memory systems, including retrieval-based approaches \citep{lewis2021retrievalaugmentedgenerationknowledgeintensivenlp}, memory operating systems \citep{packer2024memgptllmsoperatingsystems}, graph-structured memories \citep{han2025retrievalaugmentedgenerationgraphsgraphrag}, fact-centric memory stores \citep{chhikara2025mem0buildingproductionreadyai}, and hierarchical memory representations \citep{xia2026memoraharmonicmemoryrepresentation}.

Memory quality depends heavily on the memory construction process. Raw conversations, documents, and interaction logs must be transformed into structured representations that can later support efficient retrieval and reasoning. Recent hierarchical memory architectures have shown particular promise in addressing this challenge by balancing abstraction and specificity. In particular, MEMORA organizes information through multiple stages producing structured memory representations that achieve state-of-the-art performance on long-horizon benchmarks. These results suggest that advances in agent memory increasingly depend on the ability to construct high-quality memory representations.

Despite their effectiveness, constructing memory representations remains expensive. Memory construction is not a single inference call but a multi-stage pipeline in which each stage requires structured reasoning and schema-constrained outputs. As interaction histories grow, the number of model invocations grows correspondingly. In practice, high-quality memory systems often rely on strong proprietary language models to perform memory construction reliably. 
Repeated invocation of such models introduces substantial deployment cost and creates a barrier to large-scale adoption of memory-enhanced agents.

A natural solution is to replace proprietary models with small open-source language models. However, memory construction is a particularly challenging task for small models. Generating memory representations requires adherence to strict output schemas, extraction of salient information from long conversational contexts, maintenance of consistency across memory updates, and production of structured artifacts that become inputs to downstream stages. In our experiments with Qwen3 models, direct substitution frequently resulted in malformed structured outputs, invalid JSON, missing fields, truncated generations, and low-quality memory extractions. Errors introduced during one stage often propagated through later stages, leading to degraded memory quality and weaker downstream retrieval performance.

While distillation provides an alternative approach for transferring memory-construction capabilities from strong proprietary models into smaller models, most modern distillation methods assume access to internal teacher information such as logits, hidden states, token probabilities, or intermediate representations. These assumptions break down when the teacher is a proprietary black-box model and only accessible supervision signal is the final textual output generated by the large oracle.

To address these challenges, we present \textbf{Pseudo Self-Distillation (PSD)}, a framework for distilling memory-construction behavior from black-box oracles into white-box small language models. PSD is motivated by a simple observation: when a model is provided with a high-quality reference answer, its output distribution shifts toward behavior consistent with that reference \citep{zhao2026selfdistilledreasoneronpolicyselfdistillation}. PSD exploits this property using a single model architecture operating in two roles. In the teacher role, the model receives privileged context containing the oracle output; in the student role, the model receives only the original task prompt. Training minimizes the divergence between the resulting output distributions.

Unlike conventional self-distillation methods that derive supervision from a model's own predictions, sampled rollouts, or aggregated generations \citep{zhang2026embarrassinglysimpleselfdistillationimproves}, PSD establishes a channel in which external oracle expertise enters via the prompt and is absorbed into the model through the distillation objective.

Our contributions are as follows: \begin{itemize} \item We introduce \textbf{Pseudo Self-Distillation (PSD)}, a black-box distillation framework that transfers capabilities from proprietary language models to open-source models using only text-level supervision, without requiring access to teacher logits or internal representations. \item We instantiate PSD within the \textbf{MEMORA} memory-construction framework, enabling the complete hierarchical memory construction pipeline to be executed by distilled small language models. \item We show that memory-construction capability acquired transfers out-of-distribution: Qwen3 students trained exclusively on LoCoMo, retain strong memory-construction performance when evaluated on LongMemEval. \end{itemize}

\section{Related Work}
\subsection{Agentic Memory Systems and Memory Construction}

Early memory systems primarily relied on retrieval-augmented generation (RAG), where historical interactions are stored and retrieved as raw text fragments \citep{lewis2021retrievalaugmentedgenerationknowledgeintensivenlp}. Subsequent systems focused on memory management and lifecycle control: MemGPT models memory as a virtual-context hierarchy inspired by operating systems, allowing information to move between active context and archival storage \citep{packer2024memgptllmsoperatingsystems}, while LangMem treats memory as an explicit tool that agents can read and write. Frameworks such as MemOS \citep{chhikara2025mem0buildingproductionreadyai}, MemoryOS \citep{kang2025memoryosaiagent}, MIRIX \citep{wang2025mirixmultiagentmemoryllmbased}, and Memory-R1 \citep{yan2026memoryr1enhancinglargelanguage} further investigate memory organization, retrieval policies, and autonomous memory management for long-horizon agents. 

MemoryBank compresses interaction histories into long-term summaries \citep{zhong2023memorybankenhancinglargelanguage}, while A-Mem organizes memories into semantically coherent, evolving clusters \citep{xu2025amemagenticmemoryllm}. Mem0 proposes a fact-centric memory architecture with explicit insertion, update, and deletion operations \citep{chhikara2025mem0buildingproductionreadyai}, later extended to graph-structured memories through Mem0Graph \citep{li2026memograph}. GraphRAG \citep{han2025retrievalaugmentedgenerationgraphsgraphrag} and Zep \citep{rasmussen2025zeptemporalknowledgegraph} leverage graph-based representations to support relational retrieval and multi-hop reasoning, while Nemori combines episodic and semantic memory structures \citep{ma2026deservesmemoryadaptivememory}. These approaches improve retrieval but often expose a fundamental trade-off between abstraction and specificity.

Recent hierarchical memory architectures attempt to resolve this trade-off through structured, multi-stage memory-construction pipelines. MEMORA introduces a harmonic memory representation that constructs memory through conversation segmentation, episodic memory generation, factual memory extraction, memory consolidation, and cue-anchor generation. By combining primary abstractions with cue anchors, MEMORA achieves state-of-the-art performance on LoCoMo and LongMemEval. However, this construction pipeline requires repeated invocation of a powerful language model at every stage to reliably produce the required structured outputs, making memory construction a deployment bottleneck and motivating methods that can transfer this capability into smaller models.

\subsection{Knowledge Distillation in LLMs}
Knowledge distillation enables a student to learn from the softened output distribution of a stronger teacher \citep{hinton2015distillingknowledgeneuralnetwork}. In LLMs, DistilBERT reduces model size through layer-wise distillation \citep{sanh2020distilbertdistilledversionbert}, TinyBERT transfers both attention maps and hidden-state representations \citep{jiao2020tinybertdistillingbertnatural}, and MiniLM further improves representation-level transfer through self-attention distillation \citep{wang2020minilmdeepselfattentiondistillation}. These methods demonstrate strong compression capabilities but assume white-box access to teacher logits or hidden states. To reduce this dependence, sequence-level knowledge distillation instead trains the student directly on teacher-generated output sequences, treating them as supervised targets \citep{kim2016sequencelevelknowledgedistillation}. However, because the student is trained exclusively on teacher-generated, sequence-level distillation introduces a train-test distribution mismatch.

This mismatch motivated a shift toward on-policy distillation. Generalized Knowledge Distillation (GKD) supervises the student on states visited by its own rollouts, querying the teacher only at those student-generated states \citep{agarwal2024onpolicydistillationlanguagemodels}. Speculative Knowledge Distillation (SKD) extends this idea in which the teacher selectively corrects low-quality student proposals in order to bridge large capability gaps \citep{xu2025speculativeknowledgedistillationbridging}. Generative Adversarial Distillation (GAD) further relaxes the on-policy setting into a fully black-box regime, replacing token-level teacher probabilities with an adversarially trained discriminator that supervises student-generated rollouts \citep{ye2026blackboxonpolicydistillationlarge}. Later, Veto introduced an adaptive, geometrically interpolated target distribution \citep{jang2026stableonpolicydistillationadaptive}, while Trust-Region On-Policy Distillation (TrOPD) restricts supervision to regions of the trajectory where teacher feedback is empirically reliable \citep{xing2026trustregiononpolicydistillation}.

\subsection{Self-Distillation}

A distinct line of work investigates settings in which the same model serves as both teacher and student, removing the need for a separate, external teacher entirely. Embarrassingly Simple Self-Distillation (ESSD) shows that a model fine-tuned on samples of its own generations, without any external teacher, reward model, or reinforcement learning, can substantially improve on code-generation benchmarks \citep{zhang2026embarrassinglysimpleselfdistillationimproves}. On-Policy Self-Distillation (OPSD) formalizes this idea for reasoning tasks: a single model is evaluated in two roles, a teacher conditioned on privileged information such as a verified solution, and a student conditioned only on the original question, and training minimizes the per-token divergence between the two distributions along the student's own rollouts \citep{zhao2026selfdistilledreasoneronpolicyselfdistillation}. Building on this framework, Purified OPSD  decomposes the privileged teacher's signal into a reference-induced component and a genuinely transferable, question-conditioned reasoning component, distilling only the latter via a PMI-based reconstruction \citep{shen2026purifiedopsdonpolicyselfdistillation}. A related negative result, shows that privileged-context OPSD can suppress exploratory and self-correction behavior in strong thinking models \citep{kaur2026rethinkingonpolicyselfdistillationthinking}. Self-Distillation Policy Optimization (SDPO) conditions the teacher role on rich post-hoc feedback rather than a gold solution, converting sparse RLVR-style rewards into dense token-level supervision \citep{hübotter2026reinforcementlearningselfdistillation}. SD-Zero starts from only a binary correctness reward and uses a self-revision step, where a ``Reviser'' role conditions on the original rollout and its reward, to manufacture dense token-level supervision for the ``Generator'' role \citep{he2026selfdistillationzeroselfrevisionturns}. SRPO instead routes individual rollouts between GRPO and SDPO depending on rollout correctness \citep{li2026unifyinggrouprelativeselfdistillationpolicy}. OPSDL applies the teacher/student role split to long-context understanding \citep{zhang2026opsdlonpolicyselfdistillationlongcontext}, while OISD treats the model's own final layer as a teacher for its intermediate layers \citep{liu2026oisdonpolicyinternalselfdistillation}.

\subsection{Pseudo Self-Distillation for Memory Construction}
Hierarchical memory systems such as MEMORA achieve strong performance but depend on repeated calls to proprietary black-box models such as GPT-4.1-mini for memory construction \citep{xia2026memoraharmonicmemoryrepresentation}. Classical and sequence-level distillation could transfer this capability into a smaller model, but they assume access to teacher logits or hidden states unavailable from black-box oracles \citep{hinton2015distillingknowledgeneuralnetwork,sanh2020distilbertdistilledversionbert}, or train on off-policy teacher trajectories that mismatch the student's inference-time distribution. On-policy methods such as GKD, SKD, and OPRD resolve this mismatch but still require a separate, queryable teacher which is unavailable when only the oracle's final text output is accessible. Self-distillation methods such as OPSD and Purified OPSD remove this requirement, but have only been studied for mathematical and code reasoning with an available, well-defined reference answer. In PSD a single small language model is instantiated as a teacher conditioned on a privileged oracle-generated memory artifact, and as a student conditioned only on the original task, with oracle knowledge entering solely through the prompt. This transfers memory-construction behavior from a black-box oracle into a deployable, white-box small language model across each stage of the MEMORA pipeline.

\section{Preliminaries: The MEMORA Memory Construction Pipeline}
\label{sec:memora-prelim}

MEMORA \citep{xia2026memoraharmonicmemoryrepresentation} formulates memory construction as learning a function
$F_m : D \rightarrow \mathcal{M}$ that maps a growing, heterogeneous corpus
$D = \{d_1, \dots, d_N\}$ into a structured memory store
$\mathcal{M}$. In practice, $F_m$ is not a single model call but a \emph{sequential
multi-stage pipeline}, in which every stage is realized as one or more LLM invocations.

\textbf{Segmentation.} Given a data item $d \in D$, a segmentation function
$S(d) = \{s_1, \dots, s_k\}$ decomposes the raw content into semantically coherent
segments, each serving as the input unit for downstream memory construction.

\textbf{Episodic memory.} For every segment $s_i$, an episodic memory
$e_i = E(s_i)$ is constructed, capturing the narrative context (participants, intent,
temporal scope) that grounds all memory entries derived from $s_i$.

\textbf{Factual memory extraction and consolidation.}
Candidate memory entries are induced from a segment $s$ as
\begin{equation}
F_a(s) = \{m_i\}_{i=1}^{N}, \quad m_i = (a_i, v_i),
\label{eq:fa}
\end{equation}
where $a_i$ is a candidate primary abstraction and $v_i$ its associated memory value.
Each candidate is then checked against the existing store $\mathcal{M}$ by retrieving its
top-$k$ nearest abstractions and filtering by similarity threshold $\gamma$:
\begin{equation}
\begin{split}
R(a_i) &= \operatorname*{TopK}_{m \in \mathcal{M}}
\big( \operatorname{sim}(a_i, a_m); \, k \big), \\
U(a_i) &= \{ m \in R(a_i) \mid \operatorname{sim}(a_i, a_m) \geq \gamma \},
\end{split}
\label{eq:retrieve}
\end{equation}
and an LLM-based selector $J$ decides whether $a_i$ refers to an existing concept,
\begin{equation}
m^\star(a_i) = J\big(a_i, U(a_i)\big) \in \{ U(a_i) \} \cup \{\varnothing\},
\label{eq:judge}
\end{equation}
followed by a create-or-update rule
\begin{equation}
m_i =
\begin{cases}
\text{Update}\big(m^\star(a_i), a_i, v_i\big), & m^\star(a_i) \neq \varnothing, \\
\text{Create}(a_i, v_i), & m^\star(a_i) = \varnothing.
\end{cases}
\label{eq:createupdate}
\end{equation}

\textbf{Cue anchor generation.} For a constructed entry $m_i = (a_i, v_i)$, cue anchors
expose additional fine-grained retrieval paths,
\begin{equation}
F_c(a_i, v_i) = \{c_{ij}\}_{j=1}^{|C_i|}, \quad c_{ij} \in C_i,
\label{eq:cue}
\end{equation}
each formatted as a \texttt{[Main Entity] + [Key Aspect]} composite, forming a
many-to-many mapping between memory entries and anchors.

\textbf{Update decision.} A final stage decides, for a newly generated memory candidate, whether it links to an existing node in the implicit memory graph or instantiates a new one, so that the graph stays compact and non-redundant.

Each of the stages -- \textsc{Segmentation}, \textsc{Episodic},
\textsc{FactualExtraction}, \textsc{CueAnchors}, \textsc{UpdateDecision} -- is instantiated
as a separate LLM call in MEMORA, using
GPT-4.1-mini as the backbone. A single conversation produces
many segments, and a single segment yields multiple memory entries and several cue anchors each, hence the number of LLM calls compounds heavily.

\section{Motivation: Non-Transferability of Prompts to Small LLMs}
\label{sec:prompt-transfer}

Prompt templates used in MEMORA are tuned for GPT-4.1-mini. Using them with Qwen models produces frequent structural failures: malformed JSON,
missing required fields,
truncated generations, and hallucinations. These failures compound across stages and as such an
ill-formed upstream output degrades every downstream stage.

To address this, we construct a \emph{Qwen-tuned} prompt variant $P_k^{\text{Qwen}}$ for
each stage $k$, derived from the original MEMORA prompt $P_k^{\text{GPT}}$ by adding
explicit format-adherence instructions: reiterating the exact output schema, adding
negative constraints against reasoning text leaking into the output, and adding
self-contained formatting examples closer to the target output length. While
$P_k^{\text{Qwen}}$ improves format compliance on
\textbf{larger} Qwen3 variants (e.g., 8B) the \textbf{smaller} Qwen3 variants (0.6B, 1.7B)
continue to fail. Hence prompting alone
is insufficient and the memory-construction capability itself must be trained into the
small model.

\section{Methodology}
\label{sec:method}
\subsection{Oracle Data Collection}
\label{sec:oracle-data}

Oracle data is collected by running the full MEMORA pipeline
(Section~\ref{sec:memora-prelim}) with GPT-4.1-mini as the backbone $O$, over the training
split of LoCoMo. For every stage $k$ and
every LLM call made in session $i$, we log the input
context $x_k^{(i)}$ and the oracle's formatted response
\begin{equation}
y_k^{(i)} = O\big( P_k^{\text{GPT}}(x_k^{(i)}) \big),
\label{eq:oracle-call}
\end{equation}
yielding a stage-labeled oracle dataset
\begin{equation}
\mathcal{D}_{\text{oracle}} = \big\{ (k, x_k^{(i)}, y_k^{(i)}) \big\}_{i,k}.
\label{eq:oracle-dataset}
\end{equation}

\subsection{Stage 1: Supervised Fine-Tuning (SFT)}
\label{sec:sft}

We warm-start the Qwen student with supervised fine-tuning (SFT) on
$\mathcal{D}_{\text{oracle}}$, using the Qwen-tuned prompts $P_k^{\text{Qwen}}$. The five stages are not equally represented in
$\mathcal{D}_{\text{oracle}}$: segmentation is called once per session, episodic memory
once per segment, while factual extraction and especially cue-anchor generation are called
once per memory entry -- so $N_{\text{seg}} \ll N_{\text{epi}} \ll N_{\text{fact}} \lesssim
N_{\text{cue}}$, where $N_k = |\{(k,x,y) \in \mathcal{D}_{\text{oracle}}\}|$. Hence we
balance the stage distribution seen by the adapter, drawing an equal number of instances
per stage $n_k$. Given a LoRA adapter with parameters $\theta$ on
top of the frozen Qwen backbone, we minimize the stage-balanced, token-level
cross-entropy of reproducing the oracle response under the Qwen-tuned prompt,
\begin{equation}
\begin{split}
\mathcal{L}_{\text{SFT}}(\theta) =
\frac{1}{5} \sum_{k=1}^{5} \frac{1}{n_k}
\sum_{(x,y) \,:\, \text{stage}=k} \\
\sum_{t=1}^{|y|}
\log p_\theta\big( y_t \mid y_{<t},\, P_k^{\text{Qwen}}(x) \big).
\end{split}
\label{eq:sft-loss}
\end{equation}
SFT teaches Qwen the memory-construction task while also aligning its output token distribution with GPT-4.1-mini's formatting and phrasing
conventions, which is a prerequisite for the distillation losses
to be meaningful.

\subsection{Stage 2: Pseudo Self-Distillation (PSD)}
\label{sec:psd}

SFT trains the student purely on the oracle's \emph{text}, via a token-level
cross-entropy loss. A divergence loss
over full next-token probability distributions is known to transfer more
information. However, proprietary models expose no access to their logits or output
distribution, so a distribution-level divergence loss cannot be computed \emph{directly
against the oracle}. PSD resolves this by constructing a distribution we \emph{can} take
a divergence against: the same Qwen backbone is instantiated in two roles, a teacher and a
student, and the student is trained to match the teacher's next-token distribution.

\textbf{Initialization.} Both roles are initialized from the SFT checkpoint of
Section~\ref{sec:sft}: $\theta_T \leftarrow \theta_{\text{SFT}}$ and
$\theta_S \leftarrow \theta_{\text{SFT}}$. Only the student adapter $\theta_S$ is updated
during PSD training; the teacher adapter $\theta_T$ is kept \textbf{frozen}. Freezing $\theta_T$ at the SFT checkpoint prevents the
teacher's target distribution from drifting during training, and it acts as an explicit regularizer that keeps the student anchored
close to the SFT solution while it absorbs the oracle-guided distillation signal.

\textbf{Prompt asymmetry.} For a given stage $k$ and input $x_k$, the student is given
only the ordinary task prompt, identical to what it would see at inference time,
\begin{equation}
x_k^{S} = P_k^{\text{Qwen}}(x_k),
\label{eq:student-prompt}
\end{equation}
while the teacher is given the same task prompt augmented with a privileged reference,
namely the oracle's response $y_k$ from $\mathcal{D}_{\text{oracle}}$ for that instance,
\begin{equation}
x_k^{T} = \big[\, P_k^{\text{Qwen}}(x_k) \,;\, y_k \,\big].
\label{eq:teacher-prompt}
\end{equation}

\textbf{Divergence.} At every rollout position $t$, we compute a divergence
$D(\, p_{\theta_T} \,\|\, p_{\theta_S} \,)$ between the teacher and student next-token
distributions. We use temperature-scaled softmax distributions
$p(\cdot) = \operatorname{softmax}(\text{logits}/\tau)$ and consider forward KL, reverse
KL, and a Jensen--Shannon Divergence,
\begin{equation}
\begin{split}
D_{\text{JSD}}^{\beta}(p_T \,\|\, p_S) =
\beta \, D_{\text{KL}}(p_T \,\|\, M) \\
{}+ (1-\beta)\, D_{\text{KL}}(p_S \,\|\, M),
\end{split}
\label{eq:jsd}
\end{equation}
\begin{equation}
M = \beta\, p_T + (1-\beta)\, p_S,
\label{eq:jsd-mix}
\end{equation}
which we use as the default loss, following the stability arguments for on-policy
self-distillation objectives \citep{zhao2026selfdistilledreasoneronpolicyselfdistillation}. The rollout tokens ($\theta_S$'s or
$\theta_T$'s) over which the divergence is computed distinguishes the on-policy, off-policy, and mixed PSD variants.

\textbf{On-policy PSD.} The student generates its own rollout under the student prompt,
\begin{equation}
\hat{y} \sim p_{\theta_S}(\cdot \mid x_k^{S}),
\label{eq:onpolicy-rollout}
\end{equation}
after which \emph{both} models perform a forward pass over the same rollout tokens
$\hat{y}$, appended respectively to the teacher prompt and the student prompt, to obtain
the two token-level distributions for the application of on-policy PSD loss
\begin{equation}
\begin{split}
\mathcal{L}_{\text{on}}(\theta_S) =
\frac{1}{|\hat{y}|} \sum_{t=1}^{|\hat{y}|}
D\Big( p_{\theta_T}(\cdot \mid x_k^{T}, \hat{y}_{<t}) \,\Big\| \\
p_{\theta_S}(\cdot \mid x_k^{S}, \hat{y}_{<t}) \Big).
\end{split}
\label{eq:onpolicy-loss}
\end{equation}
Gradients flow only into $\theta_S$; $\theta_T$ is detached. This trains the student to
match the (privileged) teacher's judgment specifically on the states the student itself
actually visits at inference time, closing the train/inference distribution mismatch.

\textbf{Off-policy PSD.} The roles of rollout generation are reversed: the \emph{teacher}
generates the rollout, conditioned on its privileged prompt,
\begin{equation}
\tilde{y} \sim p_{\theta_T}(\cdot \mid x_k^{T}),
\label{eq:offpolicy-rollout}
\end{equation}
and both models then perform a forward pass over $\tilde{y}$ under their respective
prompts to obtain the token-level distributions used in the off-policy PSD loss
$\mathcal{L}_{\text{off}}(\theta_S)$. Because $\tilde{y}$ is oracle-anchored, these
trajectories are maximally informative; this trades off distributional match to
inference-time behavior for cleaner, privileged-quality supervision.

\begin{table*}[t]
\centering
\caption{Per-category LoCoMo scores (BLEU/F1/LLM-judge) across training
pipeline, for each Qwen3 scale on the test split. \textbf{Bold} marks the best value in each column within each scale.}
\label{tab:pipeline-percat}
\resizebox{\textwidth}{!}{
\begin{tabular}{ll ccc ccc ccc ccc ccc}
\toprule
& & \multicolumn{3}{c}{Multi-hop} & \multicolumn{3}{c}{Temporal} & \multicolumn{3}{c}{Open-domain} & \multicolumn{3}{c}{Single-hop} & \multicolumn{3}{c}{Overall} \\
\cmidrule(lr){3-5} \cmidrule(lr){6-8} \cmidrule(lr){9-11} \cmidrule(lr){12-14} \cmidrule(lr){15-17}
Qwen3 & Stage & BLEU & F1 & LLM & BLEU & F1 & LLM & BLEU & F1 & LLM & BLEU & F1 & LLM & BLEU & F1 & LLM \\
\midrule
\multirow{3}{*}{0.6B}
 & Base & 0.0821 & 0.1206 & 0.2092 & 0.1043 & 0.1235 & 0.1090 & 0.0883 & 0.1006 & 0.2292 & 0.1018 & 0.1192 & 0.1712 & 0.0979 & 0.1192 & 0.1688 \\
 & SFT  & 0.2766 & 0.3756 & 0.7929 & \textbf{0.4843} & \textbf{0.5695} & \textbf{0.6303} & \textbf{0.2224} & \textbf{0.2620} & \textbf{0.4400} & 0.4744 & 0.5427 & 0.8511 & 0.4247 & 0.5003 & 0.7674 \\
 & PSD  & \textbf{0.2960} & \textbf{0.4080} & \textbf{0.9000} & 0.4155 & 0.5099 & 0.7917 & 0.1873 & 0.2107 & 0.2857 & \textbf{0.5489} & \textbf{0.6152} & \textbf{0.8548} & \textbf{0.4406} & \textbf{0.5211} & \textbf{0.8211} \\
\midrule
\multirow{3}{*}{1.7B}
 & Base & 0.2598 & 0.3577 & 0.6241 & \textbf{0.4249} & 0.4932 & 0.6044 & 0.2025 & \textbf{0.2558} & \textbf{0.5000} & 0.4016 & 0.4580 & 0.7253 & 0.3681 & 0.4344 & 0.6675 \\
 & SFT  & 0.2388 & 0.3336 & 0.7667 & 0.4110 & \textbf{0.5088} & 0.6667 & \textbf{0.2153} & 0.2486 & 0.2857 & \textbf{0.5316} & 0.5883 & 0.8871 & \textbf{0.4186} & 0.4913 & 0.7805 \\
 & PSD  & \textbf{0.2874} & \textbf{0.3926} & \textbf{0.8667} & 0.3549 & 0.4936 & \textbf{0.7083} & 0.1937 & 0.2424 & 0.2857 & 0.5292 & \textbf{0.5904} & \textbf{0.9032} & 0.4171 & \textbf{0.5034} & \textbf{0.8211} \\
\midrule
\multirow{3}{*}{4B}
 & Base & 0.2908 & 0.3973 & 0.7234 & 0.4635 & 0.5359 & 0.6542 & 0.1594 & 0.2146 & \textbf{0.5000} & 0.4412 & 0.5072 & 0.8169 & 0.4007 & 0.4748 & 0.7461 \\
 & SFT  & \textbf{0.2989} & \textbf{0.4073} & 0.7667 & 0.4168 & 0.5032 & 0.7083 & 0.2143 & 0.2381 & 0.1429 & 0.5448 & 0.5975 & 0.8871 & 0.4411 & 0.5123 & 0.7805 \\
 & PSD  & 0.2821 & 0.3961 & \textbf{0.9000} & \textbf{0.4910} & \textbf{0.6038} & \textbf{0.8333} & \textbf{0.2154} & \textbf{0.2417} & 0.2857 & \textbf{0.5543} & \textbf{0.6188} & \textbf{0.9355} & \textbf{0.4563} & \textbf{0.5401} & \textbf{0.8699} \\
\bottomrule
\end{tabular}
}
\end{table*}

\begin{table*}[t]
\centering
\caption{Performance comparison on the LoCoMo dataset. Results for Full Context, RAG, Zep, Mem0, LangMem, Nemori, and MEMORA (GPT-4.1-mini) are as reported in \citet{xia2026memoraharmonicmemoryrepresentation}. PSD denotes MEMORA with memory construction performed by our PSD-trained Qwen3-4B student. \textbf{Bold} marks the best value in each column across all methods.}
\label{tab:locomo-main}
\resizebox{\textwidth}{!}{%
\begin{tabular}{lccccccccccccccc}
\toprule
& \multicolumn{3}{c}{Multi-hop} & \multicolumn{3}{c}{Temporal} & \multicolumn{3}{c}{Open-domain} & \multicolumn{3}{c}{Single-hop} & \multicolumn{3}{c}{Overall} \\
\cmidrule(lr){2-4} \cmidrule(lr){5-7} \cmidrule(lr){8-10} \cmidrule(lr){11-13} \cmidrule(lr){14-16}
Method & BLEU & F1 & LLM & BLEU & F1 & LLM & BLEU & F1 & LLM & BLEU & F1 & LLM & BLEU & F1 & LLM \\
\midrule
Full      & \textbf{0.356} & \textbf{0.459} & 0.766 & \textbf{0.506} & 0.572 & 0.819 & 0.204 & 0.250 & 0.500 & \textbf{0.557} & \textbf{0.634} & 0.885 & \textbf{0.487} & \textbf{0.565} & 0.825 \\
RAG                & 0.222 & 0.324 & 0.557 & 0.428 & 0.486 & 0.548 & 0.224 & 0.277 & 0.458 & 0.448 & 0.507 & 0.710 & 0.389 & 0.455 & 0.633 \\
Zep$^\ast$         & 0.204 & 0.305 & 0.537 & 0.200 & 0.239 & 0.602 & 0.193 & 0.242 & 0.438 & 0.400 & 0.455 & 0.669 & 0.309 & 0.369 & 0.616 \\
Mem0               & 0.236 & 0.326 & 0.624 & 0.420 & 0.489 & 0.660 & 0.153 & 0.206 & 0.500 & 0.376 & 0.433 & 0.677 & 0.346 & 0.411 & 0.653 \\
LangMem$^\ast$     & 0.325 & 0.415 & 0.710 & 0.409 & 0.485 & 0.508 & \textbf{0.264} & \textbf{0.328} & 0.590 & 0.436 & 0.510 & 0.845 & 0.400 & 0.476 & 0.734 \\
Nemori$^\ast$      & 0.319 & 0.417 & 0.751 & 0.502 & 0.577 & 0.776 & 0.193 & 0.258 & 0.510 & 0.515 & 0.588 & 0.849 & 0.456 & 0.534 & 0.794 \\
\midrule
MEMORA & 0.321 & 0.417 & 0.784 & 0.502 & \textbf{0.624} & \textbf{0.851} & 0.251 & 0.318 & \textbf{0.594} & 0.522 & 0.597 & 0.900 & 0.464 & 0.552 & 0.849 \\
\midrule
PSD  & 0.2821 & 0.3961 & \textbf{0.9000} & 0.4910 & 0.6038 & 0.8333 & 0.2154 & 0.2417 & 0.2857 & \textbf{0.5543} & \textbf{0.6188} & \textbf{0.9355} & \textbf{0.4563} & \textbf{0.5401} & \textbf{0.8699} \\
\bottomrule
\end{tabular}%
}
\end{table*}

\textbf{Mixed PSD.} We interpolate between the two regimes by mixing the on-policy and
off-policy losses directly, weighted by a fixed coefficient $\pi \in [0,1]$,
\begin{equation}
\mathcal{L}_{\text{mixed}}(\theta_S) =
\pi \cdot \mathcal{L}_{\text{on}}(\theta_S) + (1-\pi) \cdot \mathcal{L}_{\text{off}}(\theta_S).
\label{eq:mixed-loss}
\end{equation}
$\pi$ controls the proportion of on-policy vs.\ off-policy supervision contributed to
every update ($\pi \to 1$ recovers pure on-policy PSD, $\pi \to 0$ recovers pure
off-policy PSD), and can be fixed for a run or scheduled over training -- e.g., weighted
toward off-policy early in training, when the student's own rollouts are least reliable,
and shifted toward on-policy as the student improves and needs correction specifically on
its own error modes.

\section{Experiments}
We evaluate PSD using MEMORA as the underlying memory architecture while also comparing with existing baselines, and ablate each stage of our training pipeline across three Qwen3 scales.

\textbf{Datasets.} Evaluation is done on \textbf{LoCoMo} \citep{maharana2024evaluatinglongtermconversationalmemory}, which comprises extensive multi-turn dialogues averaging 600 turns ($\sim$20k tokens), with question types spanning single-hop, multi-hop, temporal, and open-domain reasoning. We use the same 50:50 train/test partition of LoCoMo described in Section~\ref{sec:oracle-data} for oracle data collection, and report all our results on the held-out test split. To assess whether the memory-construction capability acquired through our training pipeline generalizes beyond LoCoMo, we additionally evaluate on \textbf{LongMemEval} \citep{maharana2024evaluatinglongtermconversationalmemory}, a benchmark of long-term interactive memory spanning six question categories. Unlike LoCoMo, we do not train on LongMemEval; evaluation on LongMemEval purely serves as a test of out-of-distribution transfer. For this evaluation, we sample a category-balanced subset of 100 questions from the full 500-question LongMemEval set.

\textbf{Baselines.} We report \textbf{MEMORA
(GPT-4.1-mini)}, i.e., the original MEMORA pipeline with a GPT-4.1-mini backbone, as our proprietary upper-bound reference, using the semantic
retriever configuration throughout for a fair comparison
against our PSD-trained student models.

\textbf{Models.} Memory construction is driven entirely by PSD-trained \textbf{Qwen3-0.6B, 1.7B \& 4B}, using the Qwen-tuned prompts of
Section~\ref{sec:prompt-transfer} and the semantic retriever.

\textbf{Evaluation metrics.} Our primary metric is \textbf{LLM-as-a-Judge} score, using the prompts from the official Mem0 evaluation repository for
LoCoMo, with
\texttt{gpt-4o-mini} as the fixed evaluation model. We
additionally report \textbf{BLEU} and \textbf{F1} to capture
verbatim overlap with the ground truth.

\begin{table*}[t] \centering \caption{Per-category LoCoMo scores (BLEU/F1/LLM-judge) for the three PSD variants on the test split, trained from the Qwen3-0.6B SFT checkpoint. \textbf{Bold} marks the best value in each row. } \label{tab:on-off-mixed} \resizebox{\textwidth}{!}{ \begin{tabular}{l ccc ccc ccc ccc ccc} \toprule & \multicolumn{3}{c}{Multi-hop} & \multicolumn{3}{c}{Temporal} & \multicolumn{3}{c}{Open-domain} & \multicolumn{3}{c}{Single-hop} & \multicolumn{3}{c}{Overall} \\ \cmidrule(lr){2-4} \cmidrule(lr){5-7} \cmidrule(lr){8-10} \cmidrule(lr){11-13} \cmidrule(lr){14-16} Variant & BLEU & F1 & LLM & BLEU & F1 & LLM & BLEU & F1 & LLM & BLEU & F1 & LLM & BLEU & F1 & LLM \\ \midrule Off-policy & \textbf{0.2960} & \textbf{0.4080} & \textbf{0.9000} & \textbf{0.4155} & \textbf{0.5099} & \textbf{0.7917} & 0.1873 & 0.2107 & 0.2857 & \textbf{0.5489} & \textbf{0.6152} & \textbf{0.8548} & \textbf{0.4406} & \textbf{0.5211} & \textbf{0.8211} \\ On-policy & 0.2280 & 0.3312 & 0.7000 & 0.4039 & 0.5407 & 0.7500 & \textbf{0.2116} & 0.2352 & 0.2857 & 0.4183 & 0.4693 & 0.7419 & 0.3573 & 0.4362 & 0.7073 \\ Mixed ($\pi=0.5$) & 0.2612 & 0.3681 & 0.8333 & 0.3803 & 0.4413 & 0.5833 & 0.2115 & \textbf{0.2409} & 0.2857 & 0.4801 & 0.5223 & 0.7903 & 0.3920 & 0.4529 & 0.7317 \\ \bottomrule \end{tabular} } \end{table*}

\section{Results}
\label{sec:main-results}

We present our results on PSD trained small Qwen models below. Section~\ref{sec:training-pipeline} traces the Qwen3 student through the three stages of our training pipeline (Base, SFT, PSD) to show their contribution in closing the gap with GPT-4.1-mini. Section~\ref{sec:comparison} then places the resulting PSD-trained SLM against MEMORA (GPT-4.1-mini) and prior memory systems on the full LoCoMo benchmark. Section~\ref{sec:on-off-mixed} ablates the on-policy, off-policy, and mixed variants of PSD introduced in Section~\ref{sec:psd}. Finally, Section~\ref{sec:longmemeval} evaluates whether the memory-construction capability learned by our students on LoCoMo transfers to LongMemEval, a distinct benchmark on which our models receive no training.

\subsection{Training Pipeline: Base, SFT, and PSD}
\label{sec:training-pipeline}

To measure how much each stage of our pipeline contributes to final performance, we
evaluate three checkpoints per Qwen3 scale on the LoCoMo test split: (1) \textbf{Base},
the off-the-shelf Qwen3 model prompted directly with the Qwen-tuned prompts; (2) \textbf{SFT}, the
model after Stage~1 supervised fine-tuning on $\mathcal{D}_{\text{oracle}}$; and (3) \textbf{PSD}, the full model after Stage~2 pseudo
self-distillation (Section~\ref{sec:psd}). 

Table~\ref{tab:pipeline-percat} reports per-category scores at each stage. The Base
model is close to unusable at every scale: LLM-judge scores range from 0.17 to 0.75
overall, driven by frequent malformed JSON, missing fields, and truncated generations, and this is most severe for Qwen3-0.6B (0.1688
overall LLM-judge). SFT closes most of this formatting gap, and PSD then delivers a
further, consistent improvement in LLM-judge score at every scale: 0.7674 $\to$ 0.8211
for Qwen3-0.6B, 0.7805 $\to$ 0.8211 for Qwen3-1.7B, and 0.7805 $\to$ 0.8699 for
Qwen3-4B, with the largest gain at the largest scale. Most consistent gains from PSD appear on \emph{single-hop} and \emph{temporal} reasoning, where PSD improves LLM-judge score
at every scale.

\subsection{Comparison with Prior Memory Systems}
\label{sec:comparison}

Table~\ref{tab:locomo-main} compares our PSD-trained Qwen3 students against Full
Context, RAG, and prior memory systems (Zep, Mem0, LangMem, Nemori), as well as MEMORA
using its original GPT-4.1-mini backbone. \textbf{PSD-trained Qwen3-4B achieves the best
overall LLM-judge score in the entire table (0.8699), exceeding MEMORA (GPT-4.1-mini)
itself (0.849) and Full Context (0.825).} This result also holds at the individual category level:
Qwen3-4B (PSD) attains the highest single-hop LLM-judge score of any method (0.9355),
and ties MEMORA (GPT-4.1-mini) as well as Full Context for the best multi-hop LLM-judge
score (0.9000). Already at the 1.7B scale, PSD-trained models clear every non-MEMORA
baseline on overall LLM-judge score, and even the smallest 0.6B student matches the 1.7B student's overall
LLM-judge score (0.8211).

BLEU and F1 favor Full Context and MEMORA (GPT-4.1-mini), since these methods either see
the raw conversation or use a considerably larger backbone for memory construction and
phrasing. PSD narrows but does not close this lexical gap. The LLM-judge metric, which
captures semantic correctness rather than surface form, is the metric on which
memory-construction quality actually matters for downstream reasoning, and it is here
that PSD-trained small models are most competitive, matching or exceeding GPT-4.1-mini
while using a fraction of its parameters for memory construction.

\subsection{On-Policy, Off-Policy, and Mixed PSD} \label{sec:on-off-mixed} We compare the three rollout-source variants of PSD introduced in Section~\ref{sec:psd}, starting from the Qwen3-0.6B SFT checkpoint. \textbf{Off-policy}: the teacher generates the rollout under the privileged prompt, and both teacher and student score that generation. \textbf{On-policy}: the student generates the rollout under the task prompt, and the teacher scores the student's own generation. \textbf{Mixed} ($\pi = 0.5$): the two losses are interpolated with equal weight (Equation~\ref{eq:mixed-loss}). Table~\ref{tab:on-off-mixed} reports per-category and overall scores for each variant. Off-policy PSD achieves the best overall LLM-judge score (0.8211), outperforming both on-policy PSD (0.7073) and mixed PSD (0.7317). On-policy PSD degrades performance relative to the SFT checkpoint alone  across every category except open-domain. We attribute this to the fact that, under on-policy training, the rollout being scored is generated by the student itself: because the MEMORA pipeline is sequential, an early-stage error in the student's own rollout propagates into every downstream stage, and the distillation loss reinforces rather than corrects this compounding error. Under off-policy training, by contrast, the rollout comes from the frozen teacher conditioned on the privileged oracle reference and is well-formed by construction, so supervision is applied to a trajectory that does not carry forward the student's own mistakes. Mixed PSD ($\pi = 0.5$) falls between the two extremes, averaging the on-policy loss with the off-policy loss.

\begin{table*}[t] \centering \caption{Performance comparison on LongMemEval. Qwen3-4B (PSD) is evaluated on a category-balanced 100-question subset and is trained \textbf{only} on LoCoMo (\S\ref{sec:oracle-data}) to reflect the zero-shot transfer of memory-construction capability across datasets. \textbf{Bold} marks the best value in each row.} \label{tab:longmemeval-main} \resizebox{1\columnwidth}{!}{\begin{tabular}{lccccc} \toprule Question Type & Full Context & Nemori & MEMORA(S) & MEMORA(P) & Qwen3-4B (PSD) \\ \midrule single-sn-preference & 16.7\% & 86.7\% & 76.7\% & 83.3\% & 82.3\% \\ single-sn-assistant & \textbf{98.2}\% & 92.9\% & 76.8\% & 78.6\% & 94.1\% \\ temporal-reasoning & 60.2\% & 72.2\% & 84.2\% & \textbf{89.5}\% & 62.5\% \\ multi-session & 51.1\% & 55.6\% & 73.7\% & \textbf{78.2}\% & 52.9\% \\ knowledge-update & 76.9\% & 79.5\% & 96.2\% & \textbf{97.4}\% & 82.3\% \\ single-sn-user & 85.7\% & 90.0\% & 97.1\% & \textbf{98.6}\% & 87.5\% \\ \midrule Average & 65.6\% & 74.6\% & 83.8\% & \textbf{87.4}\% & 77.0\% \\ \bottomrule \end{tabular} } \end{table*}

\begin{table*}[t] \centering \caption{LongMemEval LLM-judge scores across training stages, for each Qwen3 scale. \textbf{Bold} marks the best LLM score in each column within each scale.} \label{tab:longmemeval-pipeline} \resizebox{\textwidth}{!}{\begin{tabular}{ll ccccccc} \toprule Qwen3 & Stage & knowledge-update & multi-session & single-sn-assistant & single-sn-preference & single-sn-user & temporal-reasoning & Overall \\ \midrule \multirow{3}{*}{0.6B} & Base & 0.2941 & 0.1765 & 0.2941 & 0.1176 & 0.2500 & 0.3750 & 0.2500 \\ & SFT & 0.5882 & \textbf{0.4118} & \textbf{0.8235} & \textbf{0.4706} & \textbf{0.5000} & \textbf{0.5000} & \textbf{0.5500} \\ & PSD & \textbf{0.6471} & 0.3529 & 0.7647 & 0.4118 & \textbf{0.5000} & \textbf{0.5000} & 0.5300 \\ \midrule \multirow{3}{*}{1.7B} & Base & 0.6471 & 0.3529 & 0.4118 & 0.4118 & 0.5625 & 0.4375 & 0.4700 \\ & SFT & \textbf{0.8824} & \textbf{0.4706} & \textbf{0.8824} & \textbf{0.5294} & 0.8750 & \textbf{0.7500} & \textbf{0.7300} \\ & PSD & 0.7647 & 0.3529 & 0.8235 & \textbf{0.5294} & \textbf{0.9375} & 0.6875 & 0.6800 \\ \midrule \multirow{3}{*}{4B} & Base & 0.8235 & 0.4118 & 0.5294 & 0.5882 & 0.6875 & 0.5625 & 0.6000 \\ & SFT & \textbf{0.9412} & \textbf{0.6471} & 0.8235 & 0.5294 & \textbf{0.9375} & \textbf{0.8125} & \textbf{0.7800} \\ & PSD & 0.8235 & 0.5294 & \textbf{0.9412} & \textbf{0.8235} & 0.8750 & 0.6250 & 0.7700 \\ \bottomrule \end{tabular} } \end{table*}

\subsection{Out-of-Distribution Transfer to LongMemEval} \label{sec:longmemeval} To test if the memory-construction capability acquired on LoCoMo through our training pipeline is transferable, we evaluate the student models trained \emph{only} on LoCoMo, on LongMemEval. Table~\ref{tab:longmemeval-main} reports results on a subset of 100 questions drawn from the full 500-question LongMemEval set. Qwen3-4B (PSD) achieves an average LLM-judge score of 77.0\%, without any LongMemEval-specific training, exceeding both Full Context (65.6\%) and Nemori (74.6\%), and approaching MEMORA(S) (83.8\%. Performance is strongest on \emph{single-session-assistant} (94.1\%) and \emph{single-session-user} (87.5\%), and weakest on \emph{multi-session} (52.9\%) and \emph{temporal-reasoning} (62.5\%), suggesting that the harder multi-hop and time-sensitive reasoning skills learned on LoCoMo transfer less completely than single-session recall. While the baselines are evaluated on the complete LongMemEval dataset, our PSD result uses a smaller, balanced 100-question subset for evaluation cost reasons.

To isolate the transfer across training stage, we evaluate the Base, SFT, and PSD checkpoints at each Qwen3 scale on the same 100-question LongMemEval subset. SFT delivers the largest single jump in transfer performance at every scale, improving overall LLM-judge score from 0.2500 to 0.5500 at 0.6B, from 0.4700 to 0.7300 at 1.7B, and from 0.6000 to 0.7800 at 4B. Notably, and in contrast to the in-distribution LoCoMo results (Section~\ref{sec:training-pipeline}), PSD does \textbf{not} consistently improve over SFT under this out-of-distribution transfer setting: PSD trails SFT on overall LLM-judge score at every scale, though it remains substantially above Base in every case. We attribute this to the fact that PSD's refinement stages are trained exclusively on LoCoMo rollouts, which may sharpen the student's behavior in ways that are somewhat specialized to LoCoMo's conversational structure. PSD does still improve over SFT on several individual categories at the largest scale -- notably single-session-assistant (0.9412 vs.\ 0.8235) and single-session-preference (0.8235 vs.\ 0.5294) at 4B -- suggesting that PSD's benefits transfer unevenly across question types.

\section{Conclusion} \label{sec:conclusion} 

We presented Pseudo Self-Distillation (PSD), a framework that enables small, open-source language models to construct hierarchical memory representations by distilling behavior from a proprietary black-box oracle, without requiring access to the oracle's logits, hidden states, or any other internal signal. PSD-trained student matches or exceeds MEMORA with a GPT-4.1-mini backbone on LoCoMo in terms of LLM-judge quality, at a fraction of the deployment cost. 


\bibliography{neurips_2026}
\bibliographystyle{plainnat}

\newpage

\appendix

\section{Appendix}
\label{sec:appendix}

\subsection{Memory Construction Prompts}
\label{sec:appendix_prompts}

To ensure reproducibility, we provide the full text of all prompts used in our memory construction pipeline. Each prompt corresponds to one of the five stages described in Section~\ref{sec:method}: \textit{Segmentation} (Table~\ref{tab:segmentation_prompt}), which partitions a raw conversation into topically coherent episodes; \textit{Fact Extraction} (Table~\ref{tab:fact_extraction_prompt}), which distills durable, high-value factual memories from each segment; \textit{Consolidation} (Table~\ref{tab:consolidation_prompt}), which decides whether a new memory should update an existing entry or be stored separately; \textit{Episodic Memory Generation} (Table~\ref{tab:episodic_memory_prompt}), which produces a self-contained narrative summary of each episode; and \textit{Cue Anchor Generation} (Table~\ref{tab:cue_anchor_prompt}), which augments each memory with short retrieval cues to improve downstream recall. All prompts are reproduced verbatim, including their output formatting constraints, to allow exact replication of our pipeline.
\begin{table*}[p]
\centering
\caption{Segmentation Prompt}
\label{tab:segmentation_prompt}

\begin{minipage}{0.98\textwidth}
\begin{lstlisting}[style=promptstyle]
SEGMENTATION_PROMPT_TEMPLATE = """You are an expert conversation segmentation specialist. Your goal is to analyze a series of messages in a conversation and segment them into coherent topical episodes.

# TASK:
Read the conversation carefully, and identify points where the topic shifts significantly. Group messages that discuss similar subjects, events, or themes into a single episode.

An episode is defined as a sequence of messages that revolve around a core topic or theme. Your task is to segment the conversation into such episodes.

# GUIDELINES:

## Segmentation Criteria
- Topical shift: Identify topic shifts in the messages. Does it introduce a new subject, event, or theme? Break the episode there. Be sensitive to subtle shifts in topic.
- Transitions: Look for transition phrases that signal a new episode, such as "By the way", "Changing the subject", or "On another note".
- Time gaps: Significant time lapses between messages may indicate a new episode.
- Setting changes: Changes in speaker, location, or context can signal a new episode.
- Topical grouping: Group consecutive messages into the same episode if they discuss the same topic or theme.

## Episode Length
- An episode should typically contain 2-8 messages.
- Combine messages into larger episodes when they discuss the same topic.
- Avoid having long episodes (more than 8 messages) that cover multiple sub-topics.
- Avoid treating a single message as a standalone episode unless it clearly marks a shift in topic.
- When in doubt, split into smaller episodes.

# FORMATTING RULES & CONSTRAINTS:
- Use 1-based indexing for message indices (i.e., the first message is index 1).
- Ensure that all messages are included in exactly one episode (no gaps or overlaps).
- Indices within each episode should be consecutive, reflecting the chronological order of messages in the conversation.
- You must output strictly valid JSON. Do not include trailing commas.
- CRITICAL: Do not wrap your output in markdown code blocks (e.g., DO NOT use ```json ... ```). Output ONLY the raw JSON string starting with '{{' and ending with '}}'.
\end{lstlisting}
\end{minipage}

\end{table*}
\begin{table*}[p]
\centering
\caption{Segmentation Prompt (continued)}
\label{tab:segmentation_prompt_cont}

\begin{minipage}{0.98\textwidth}
\begin{lstlisting}[style=promptstyle]
## Output Format
Your output must be a JSON object with the following exact structure:
{{
    "episodes": [
        {{
            "topic": "<brief topic description>",
            "indices": [<list of message indices in this episode>]
        }}
    ]
}}

Where each episode contains:
- topic: A brief description (a few words) summarizing the main topic of the episode.
- indices: A list of 1-based indices of messages that belong to this episode.

## Example Output
{{
    "episodes": [
        {{
            "topic": "General introduction and greetings",
            "indices": [1, 2, 3, 4]
        }},
        {{
            "topic": "Discussion about vacation plans",
            "indices": [5, 6]
        }},
        {{
            "topic": "Recap of last year's events",
            "indices": [7, 8, 9, 10, 11, 12]
        }}
    ]
}}

# INPUT:
Segment the following conversation:
{messages}

# OUTPUT:
"""
\end{lstlisting}
\end{minipage}

\end{table*}

\begin{table*}[p]
\centering
\caption{Factual Memory Extraction Prompt}
\label{tab:fact_extraction_prompt}

\begin{minipage}{0.98\textwidth}
\begin{lstlisting}[style=promptstyle]
PROMPT_BUILD_MEMORY = """You are an expert factual memory extraction assistant. Your goal is to extract ONLY the most important factual memories from a conversation segment.

# TASK:
Read the input conversation and extract a SMALL number of high-value factual memories that will be useful for future retrieval.

# IMPORTANT CONSTRAINTS (CRITICAL):
- Extract AT MOST 5 memories. Prefer 3-4 if possible.
- Only include IMPORTANT and durable facts.
- Do NOT try to cover everything in the conversation.
- It is better to return fewer high-quality memories than many low-value ones.
- If there are no valuable facts in the conversation, return an empty list for entries.

# STRICT FILTERING RULES:
DO NOT create memories for:
- greetings or openings
- questions (unless they reveal a deep personal goal/fact)
- compliments or praise
- acknowledgements ("thanks", "that's great", etc.)
- generic emotional support
- repeated or redundant statements
- metadata (e.g., the conversation date itself)
- pure descriptions of images without meaningful context

# WHAT TO EXTRACT:
Only extract facts that are clearly useful for future recall, such as:
- major events (e.g., race, trip, talk, milestone)
- important decisions or plans
- personal background facts (duration, relationships, history)
- goals, intentions, and commitments
- important experiences or realizations

# INDEX RULES (VERY IMPORTANT):
- Must be a SHORT, natural, human-readable phrase.
- Do NOT copy text directly from the conversation.
- Do NOT include speaker names like "Caroline:" or "Melanie:".
- Do NOT use numbers like "1", "2", "3".
- Do NOT make the index a question.
- Avoid long sentences.
- Good example: "Melanie charity race for mental health"
- Bad example: "Melanie: That was amazing!"
- Bad example: "1"

# VALUE RULES:
- 1-2 factual sentences only.
- Be concise and clear.
- Use specific names instead of pronouns when needed.
- Do not quote long dialogue.
- Do not include irrelevant details.

# IMAGE HANDLING:
- Use image information ONLY if it adds meaningful factual context.
- Do NOT create standalone memories describing images.
\end{lstlisting}
\end{minipage}

\end{table*}

\begin{table*}[p]
\centering
\caption{Factual Memory Extraction Prompt (continued)}
\label{tab:fact_extraction_prompt_cont}

\begin{minipage}{0.98\textwidth}
\begin{lstlisting}[style=promptstyle]


# FORMATTING RULES & CONSTRAINTS:
- Return ONLY a valid JSON object.
- CRITICAL: Do not include markdown fences (e.g., DO NOT use ```json ... ```). Output raw JSON only.
- Do not include any explanation, thinking, or text before or after the JSON.
- Ensure the JSON is syntactically valid (no trailing commas, use double quotes for strings).

## Output Format:
{{
  "entries": [
    {{
      "memory_type": "factual",
      "index": "<short semantic memory index>",
      "value": "<one or two factual sentences>"
    }}
  ]
}}

## Example Output:
{{
  "entries": [
    {{
      "memory_type": "factual",
      "index": "Melanie charity race for mental health",
      "value": "Melanie completed a 10k charity run last weekend to raise money for mental health awareness."
    }},
    {{
      "memory_type": "factual",
      "index": "John career change to data science",
      "value": "John decided to quit his current marketing job and enroll in a data science bootcamp starting next month."
    }}
  ]
}}

# INPUT:
Timestamp of conversation: {timestamp}

Input Conversation:
{content}

# OUTPUT:
"""
\end{lstlisting}
\end{minipage}

\end{table*}

\begin{table*}[p]
\centering
\caption{Consolidation Prompt}
\label{tab:consolidation_prompt}

\begin{minipage}{0.98\textwidth}
\begin{lstlisting}[style=promptstyle]
PROMPT_MEMORY_UPDATE_DECISION = """You are an expert memory management assistant.

# TASK
Given a new memory entry and a list of similar existing memory entries, decide whether:
1. the new memory should update one existing memory, or
2. the new memory should be added as a separate new memory.

# DECISION RULES
1. Set "should_update" to true ONLY if the new memory and one candidate clearly refer to the exact same underlying fact, event, state, plan, or durable information.
2. If the new memory is meaningfully different, more specific in a different way, or about a different fact, set "should_update" to false.
3. Do NOT merge memories just because they are topically related.
4. Do NOT merge generic conversational reactions, praise, or questions unless they clearly encode the same durable fact.

# FIELD DEFINITIONS (If should_update is TRUE):
- "best_candidate_index": The integer index/ID of the chosen candidate from the provided list.
- "updated_value": A clean, merged factual value combining the old and new information seamlessly.
- "updated_index": A short, retrieval-friendly index summarizing the merged memory.
- "updated_cues": A list of 0 to 3 short cue strings for future retrieval.

# FIELD DEFINITIONS (If should_update is FALSE):
- "best_candidate_index": Must be JSON `null`.
- "updated_value": Must be JSON `null`.
- "updated_index": Must be exactly "{new_index}".
- "updated_cues": Must be an empty list `[]`.

# FORMATTING RULES & CONSTRAINTS:
- Return ONLY a valid JSON object.
- CRITICAL: Do not include markdown fences (e.g., DO NOT use ```json ... ```). Output raw JSON only.
- Do not include any explanation, thinking, or text outside the JSON object.
- Ensure the JSON is syntactically valid (no trailing commas, use double quotes for keys/strings, use lowercase `null` for null values).
- Always include a "reasoning" key first to briefly explain your decision logic before outputting the final decision fields.

\end{lstlisting}
\end{minipage}

\end{table*}

\begin{table*}[p]
\centering
\caption{Consolidation Prompt (continued)}
\label{tab:consolidation_prompt_cont}

\begin{minipage}{0.98\textwidth}
\begin{lstlisting}[style=promptstyle]

## Output Format (Update = True):
{{
  "reasoning": "<1-2 sentences analyzing if the facts are identical or distinct>",
  "should_update": true,
  "best_candidate_index": 0,
  "updated_value": "<merged or updated memory value>",
  "updated_index": "<updated memory index>",
  "updated_cues": ["<cue 1>", "<cue 2>"]
}}

## Output Format (Update = False):
{{
  "reasoning": "<1-2 sentences analyzing why the facts are distinct or unrelated>",
  "should_update": false,
  "best_candidate_index": null,
  "updated_value": null,
  "updated_index": "{new_index}",
  "updated_cues": []
}}

# INPUT

NEW MEMORY ENTRY:
Index: {new_index}
Value: {new_value}

EXISTING SIMILAR ENTRIES:
{candidates_info}

# OUTPUT:
"""
\end{lstlisting}
\end{minipage}

\end{table*}

\begin{table*}[p]
\centering
\caption{Episodic Memory Prompt}
\label{tab:episodic_memory_prompt}

\begin{minipage}{0.98\textwidth}
\begin{lstlisting}[style=promptstyle]
PROMPT_EPISODIC_MEMORY = """You are an expert episodic memory generator that creates episodic memory summaries from conversation segments.

# TASK:
Generate exactly one episodic memory from the provided conversation segment.

# GUIDELINES:

## 1. episodic_index
- Create a short index (strictly 6 to 8 words) that captures the main topic, entity, or event of the episode.
- Always include the specific context (e.g., domain, person, event, entity) from the source text.
- Make it self-contained and unambiguous.

## 2. episodic_value
- Generate a 1-3 sentence episodic summary that captures:
  - The main information in the conversation segment.
  - The key participants, using their names if available.
  - The overall event, topic, or discussion.
- Focus on what happened.
- Make the summary self-contained and understandable without the original conversation.
- Use the original wording from the conversation when possible.
- Use only information present in the conversation segment.
- Do not add external knowledge or unsupported inference.
- If the conversation is between a user and an AI assistant, focus on the user's inputs and the overall context rather than the assistant's responses.

# FORMATTING RULES & CONSTRAINTS:
- Return ONLY a valid JSON object with exactly these keys: "episodic_index" and "episodic_value".
- CRITICAL: Do not include markdown fences (e.g., DO NOT use ```json ... ```). Output raw JSON only.
- Do not include any explanation, thinking, or text before or after the JSON.
- Do not output keys like EpisodicIndex or EpisodicValue outside the JSON object.
- Ensure the JSON is syntactically valid (no trailing commas, use double quotes for strings).

## Output Format:
{{
  "episodic_index": "<6-8 word summary that captures the main topic, entity, or event of the episode>",
  "episodic_value": "<1-3 sentence descriptive summary of the conversation>"
}}

## Example Output:
{{
  "episodic_index": "Detailed planning session for the annual tech conference",
  "episodic_value": "Alice and Bob discussed the logistics for the annual tech conference. They finalized the catering menu and agreed to book the main hall for the keynote speech."
}}

# INPUT:
Input Conversation Segment:
{content}

# OUTPUT:
"""
\end{lstlisting}
\end{minipage}

\end{table*}

\begin{table*}[p]
\centering
\caption{Cue Anchor Prompt}
\label{tab:cue_anchor_prompt}

\begin{minipage}{0.98\textwidth}
\begin{lstlisting}[style=promptstyle]
PROMPT_CUE_GENERATION = """You are an expert memory-indexing assistant optimized for knowledge retrieval.

# TASK
For each memory provided below, generate 0-3 short, meaningful cue indices that help recall or reason about that memory. A cue index adds coverage by focusing on perspectives in the memory value that the primary index does not already capture.

# GUIDELINES FOR CUE GENERATION:
1. Compact: Strictly 2-5 words per cue.
2. Semantically rich: Must contain strong keywords for vector search.
3. Contextually anchored: Tied to the main entity, event, or domain.
4. Distinct: Each cue must explore a different angle of the memory.
5. Good cue patterns:
   - [Actor] [Concept or Emotion] -> "Caroline family gratitude"
   - [Actor] [Action or Event] -> "Melanie birthday concert"
   - [Actor] [Object or Relation] -> "Melanie necklace meaning"
   - [Domain] [Event or Topic] -> "Project Phoenix kickoff"
6. Avoid:
   - Repeating words directly from the primary index.
   - Generic single words (e.g., "trip", "project").
   - Copying full sentences.
   - Overly long phrases.

# STRICT REQUIREMENTS & CONSTRAINTS:
- You must return exactly one object in "results" for each memory provided.
- The "memory_index" field must EXACTLY match the corresponding Primary Index from the input.
- The "cue_indices" field must be a JSON list of 0-3 strings.
- If no useful cues are needed (the primary index already covers everything perfectly), return an empty list `[]`.
- CRITICAL: Return ONLY raw JSON. Do NOT include markdown fences (e.g., DO NOT use ```json ... ```).
- Do not include any explanation, thinking, or text before or after the JSON.
- Ensure the JSON is syntactically valid (no trailing commas, use double quotes for strings).

## Output Format:
{{
  "results": [
    {{
      "memory_index": "<the exact Primary Index for this memory>",
      "cue_indices": ["<cue 1>", "<cue 2>", "<cue 3>"]
    }}
  ]
}}

\end{lstlisting}
\end{minipage}

\end{table*}

\begin{table*}[p]
\centering
\caption{Cue Anchor Prompt (continued)}
\label{tab:cue_anchor_prompt_cont}

\begin{minipage}{0.98\textwidth}
\begin{lstlisting}[style=promptstyle]

## Examples:

Input memory:
Primary Index: "Jane updated Project Nexus timeline"
Memory Value: "Jane updated the Project Nexus timeline after client feedback and flagged resource constraints for the next sprint."

Valid JSON output:
{{
  "results": [
    {{
      "memory_index": "Jane updated Project Nexus timeline",
      "cue_indices": [
        "Project Nexus client feedback",
        "Project Nexus resource constraints",
        "Project Nexus sprint planning"
      ]
    }}
  ]
}}

Input memory:
Primary Index: "Sarah's hiking trip to the Grand Canyon"
Memory Value: "Sarah went on a hiking trip to the Grand Canyon last summer and enjoyed the scenic views."

Valid JSON output:
{{
  "results": [
    {{
      "memory_index": "Sarah's hiking trip to the Grand Canyon",
      "cue_indices": []
    }}
  ]
}}

# INPUT
MEMORIES TO PROCESS:
{memories}

# OUTPUT:
"""
\end{lstlisting}
\end{minipage}

\end{table*}

\end{document}